\documentclass[aps,superscriptaddress,prc,twocolumn,nofootinbib]{revtex4}

\usepackage{amsmath,bm}
\usepackage{graphicx}
\usepackage{epstopdf}
\usepackage{subfigure}
\usepackage{epsfig}
\usepackage{amsmath,amssymb,amsfonts}
\usepackage{color}
\usepackage[utf8]{inputenc}
\usepackage{verbatim}
\usepackage{array}
\begin{document}

\renewcommand{\topfraction}{0.85}
\renewcommand{\textfraction}{0.1}
\renewcommand{\floatpagefraction}{0.75}

\title{The global geometrical property of jet events in high-energy nuclear collisions}

\date{\today  \hspace{1ex}}
\author{Shi-Yong Chen}
\affiliation{Key Laboratory of Quark \& Lepton Physics (MOE) and Institute of Particle Physics,
 Central China Normal University, Wuhan 430079, China}

\author{Wei Dai \footnote{weidai@cug.edu.cn}}
\affiliation{School of Mathematics and Physics, China University
of Geosciences, Wuhan 430074, China}

\author{Shan-Liang Zhang}
\affiliation{Key Laboratory of Quark \& Lepton Physics (MOE) and Institute of Particle Physics,
 Central China Normal University, Wuhan 430079, China}

\author{Qing Zhang}
\affiliation{Key Laboratory of Quark \& Lepton Physics (MOE) and Institute of Particle Physics,
 Central China Normal University, Wuhan 430079, China}

\author{Ben-Wei Zhang \footnote{bwzhang@mail.ccnu.edu.cn}}
\affiliation{Key Laboratory of Quark \& Lepton Physics (MOE) and Institute of Particle Physics,
 Central China Normal University, Wuhan 430079, China}
\affiliation{Institute of Quantum Matter, South China Normal University, Guangzhou 510006, China}

\begin{abstract}
We present the first theoretical study of medium modifications of the global geometrical pattern, i.e., transverse sphericity ($S_{\perp}$) distribution of jet events with parton energy loss in relativistic heavy-ion collisions. In our investigation, POWHEG+PYTHIA is employed to make an accurate description of transverse sphericity in the p+p baseline, which combines the next-to-leading order (NLO) pQCD calculations with the matched parton shower (PS). The Linear Boltzmann Transport (LBT) model of the parton energy loss is implemented to simulate the in-medium evolution of jets. We calculate the event normalized  transverse sphericity distribution in central Pb+Pb collisions at the LHC, and give its medium modifications.
An enhancement of transverse sphericity distribution at small $S_{\perp}$ region but a suppression at large $S_{\perp}$ region are observed in A+A collisions as compared to their p+p references, which indicates that in overall the geometry of jet events in Pb+Pb  becomes more pencil-like. We demonstrate that for events with 2 jets in the final-state of heavy-ion collisions, the jet quenching makes the geometry more sphere-like with medium-induced gluon radiation. However, for events with $\ge 3$ jets, parton energy loss in the QCD medium leads to the events more pencil-like due to jet number reduction, where less energetic jets may lose their energies and then fall off the jet selection kinematic cut. These two effects offset each other and in the end result in more jetty events in heavy-ion collisions relative to that in p+p.

\end{abstract}


\maketitle

\section{introduction}
\label{sec:Intro}

Heavy ion collision experiments performed at the RHIC and the LHC are designed to study the properties of the de-confined Quark Gluon Plasma (QGP) which created shortly after these collisions~\cite{Adler:2003qi, Adams:2003kv, Adare:2008qa, Agakishiev:2011dc, Aad:2010bu, Aamodt:2010jd, Chatrchyan:2011sx}. Energetic partons produced at the initial collision will traverse through this hot and dense medium and lose their energies by interacting with such medium, it is referred as jet quenching effect~\cite{Wang:1991xy, Gyulassy:2003mc, Qin:2015srf}. This phenomenon can be quantified by various observables, from single hadron production suppression $R^h_{\rm AA}$~\cite{Chen:2010te,Chen:2011vt,Liu:2015vna,Dai:2015dxa,Dai:2017piq,Dai:2017tuy,Ma:2018swx,Xie:2019oxg} to the observables within the productions of full jets such as inclusive jets $R^{jet}_{\rm AA}$, dijets momentum imbalance $A_J$, and tagged jets momentum imbalance $z_J$, the angular correlations of the leading two jets $\Delta \Phi_{12}$ etc. The available of these full jets observables are due to the improvement of jet finding algorithm and jet reconstruction in the final state of the heavy ion collisions at the LHC. The study of these observables are mainly focused on the medium modifications of individuals or the leading two jets in the final state of collision events~\cite{Vitev:2008rz, Vitev:2009rd, CasalderreySolana:2010eh, Young:2011qx, He:2011pd, ColemanSmith:2012vr, Neufeld:2010fj, Zapp:2012ak, Dai:2012am, Ma:2013pha, Senzel:2013dta, Casalderrey-Solana:2014bpa, Milhano:2015mng, Chang:2016gjp, Majumder:2014gda, Chen:2016cof, Chien:2016led, Apolinario:2017qay, Connors:2017ptx, Luo:2018pto, Zhang:2018urd, Zhang:2018ydt, Dai:2018mhw, Wang:2019xey,Chen:2019gqo,Yan:2020cxy}.  It is of great interest to investigate the impact of the jet quenching effect to the whole picture of produced events with all the reconstructed jets in them. For such investigation, observables that can characterize the global geometrical properties of the produced events are required, and the medium alteration of these observables in A+A collisions might give insights into the mechanism of the jet quenching phenomenon or even provide further constraint on jet quenching modeling.

Event shape observables, named as thrust, sphericity, have long been proposed to study geometrical properties and patterns of the energy flow of the collisions, and thus provide a probe of multi-jet topologies in an interaction~\cite{Abbiendi:2004qz, Aktas:2005tz, Chekanov:2006hv, Bethke:2008hf, Kluth:2000km, Khachatryan:2011dx, Aaltonen:2011et, Banfi:2004nk, Banfi:2010xy, Kluth:2009nx, Becher:2015lmy, Aad:2012fza, Aad:2012np, Sirunyan:2018adt}. For example, the sphericity of an event was firstly proposed to confirm the jet hypothesis for hadron production in electron-positron collisions by G. Hanson and his collaborators at 1975~\cite{Hanson:1975fe}. Recently, experimentalists show interests of the event shape obsverables at large momentum transfer, ATLAS Collaboration at the LHC has measured the production distribution of the transverse sphericity ($S_{\perp}$) in p+p collisions at $\sqrt{s}=7$~TeV~\cite{Aad:2012np}. The larger transverse sphericity $S_{\perp}$ is, the more isotropic an event is.

In this manuscript, for the first time, we calculate the medium modification of the global geometrical property, i.e. the transverse sphericity distribution in Pb+Pb collisions at the LHC, by including jet quenching effect in the QGP. We found the event normalized $S_{\perp}$ distributions in Pb+Pb collisions are enhanced at very small $S_{\perp}$ region while suppressed at larger $S_{\perp}$ compared to the distribution in p+p reference, indicating more proportion of the survived events are shifted to pencile-like region (small $S_{\perp}$) due to the jet quenching effect. It seems however counterintuitive.  To understand such a result, we first studied the nuclear modifications of $S_{\perp}$ distributions for events with identical numbers of jets. Next, the reduction of jet numbers in events due to the jet energy loss effect during the in-medium evolution are therefore investigated. It happens when jets are discarded because their $p_{T}$ fall below the lower threshold of the jet selection.  A similar feature has also been pointed out in a study of di-jet momentum imbalance in Pb+Pb collisions at the LHC~\cite{Apolinario:2012cg}.
We further investigate the medium modification of the azimuth angle correlation ($\Delta\phi$) and transverse momentum balance ($x_{J}$) among the jets in events to get a deeper understanding on how relevant is the medium modification of the geometric structure of the events to understand the medium modification of the $S_{\perp}$ distribution.

The paper is organized as follows. We will introduce the setups of the p+p events generation and also the theoretical framework to describe the parton in-medium evolution in Sec.~\ref{sec:framework}. After that, in Section~\ref{sec:results}, the theoretical predictions of the event normalized $S_{\perp}$ distribution in p+p and its modification in Pb+Pb collisions at the LHC are presented. We further explore the nature of such modification and also present both the $\Delta\phi$ and $x_{J}$ correlations with $S_{\perp}$ for two jets events in p+p and Pb+Pb collisions. A summary will be given in Sec.~\ref{sec:summary}.

\section{p+p events generation and parton in-medium evolution}
\label{sec:framework}


The transverse sphericity is one of event shape observables defined in terms of all jets in the event,
as follows~\cite{Hanson:1975fe,Aad:2012np}:
\begin{eqnarray}
S_{\perp} = \frac{2\lambda_{2}}{\lambda_{1}+\lambda_{2}}
\label{eq:S}
\end{eqnarray}
where $\lambda_{1}$, $\lambda_{2}$ and $\lambda_{3}$ $(\lambda_{1}>\lambda_{2}>\lambda_{3})$ are the normalized individual eigenvalues of this momentum matrix:
\begin{eqnarray}
M=\sum_{i}\left(
  \begin{array}{ccc}
    p_{xi}^{2} & p_{xi}p_{yi} & p_{xi}p_{zi} \\
    p_{yi}p_{xi} & p_{yi}^{2} & p_{yi}p_{zi} \\
    p_{zi}p_{xi} & p_{zi}p_{yi} & p_{zi}^{2} \\
  \end{array}
\right) \,  ,
\label{eq:matrix}
\end{eqnarray}
where $i$ represent the $i$th jet in the event. The event is pencil-like when transverse sphericity $S_{\perp}\rightarrow 0$, while the event is sphere-like when $S_{\perp}\rightarrow 1$.

\begin{figure}[!htb]
\centering
\includegraphics[width=6cm,height=6cm]{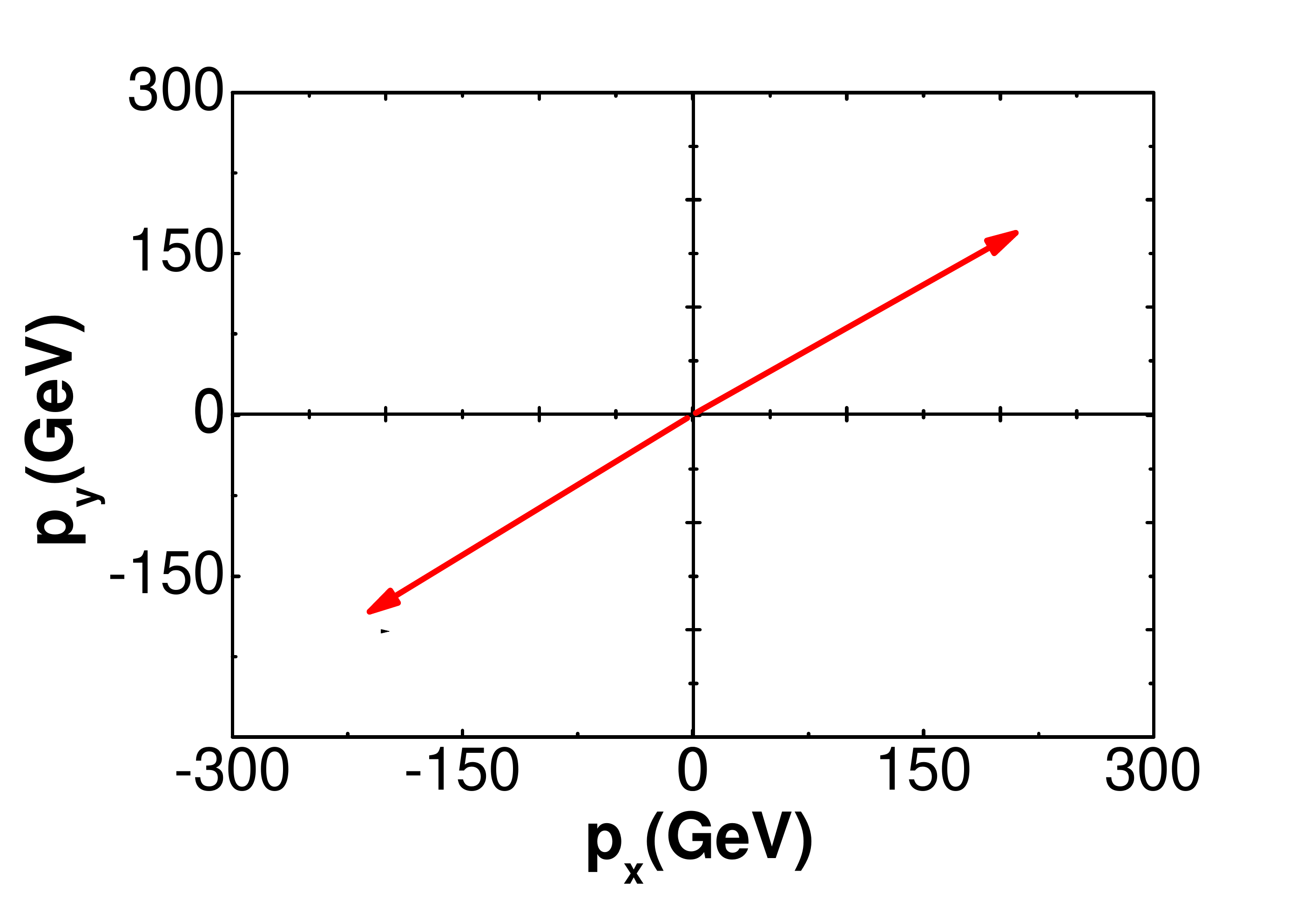}
\includegraphics[width=6cm,height=6cm]{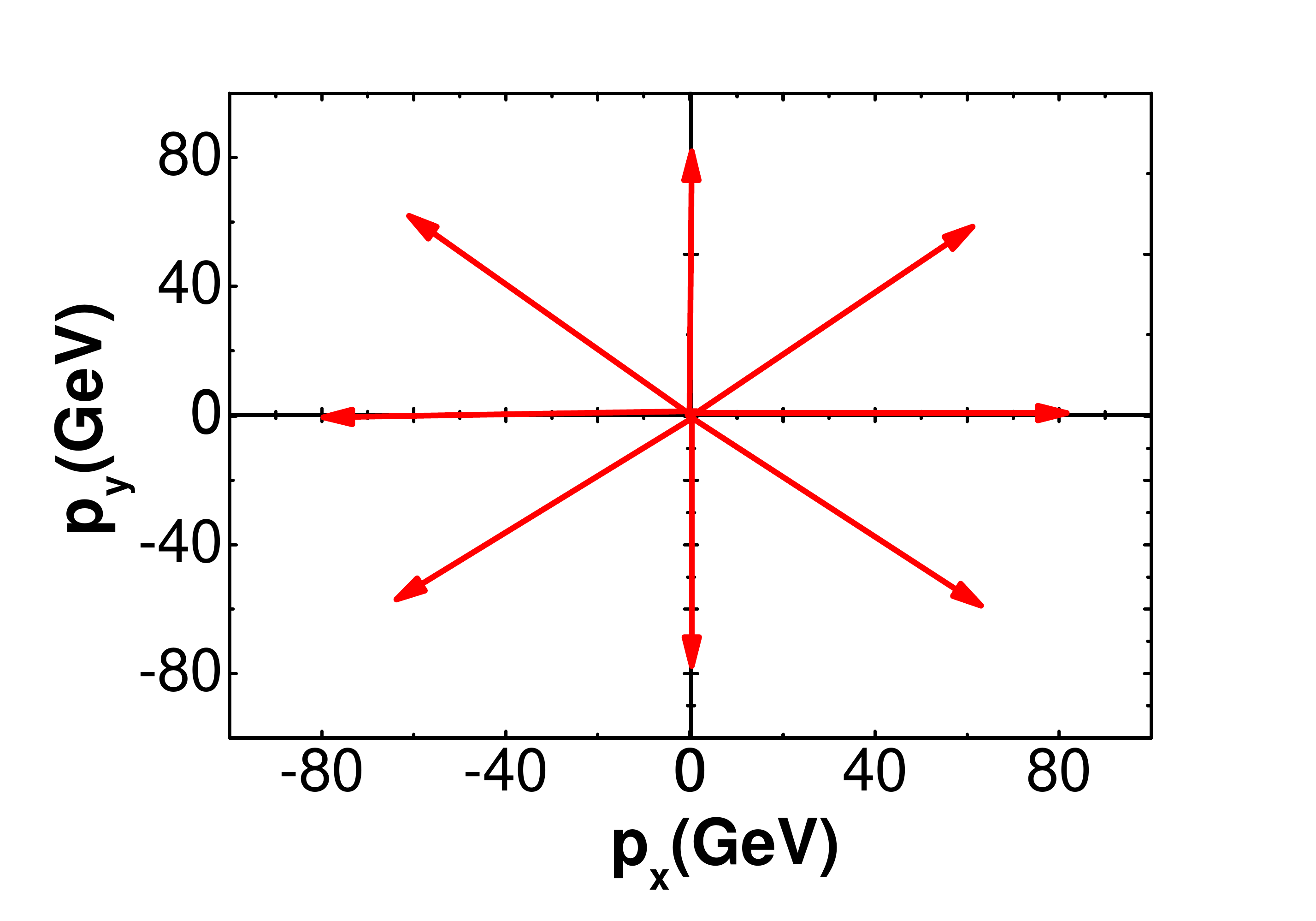}
\caption{Top: typical Pencile-like selected event with $S_\perp=0$;
Bottom: typical Sphere-like selected event with $S_\perp=1$.}
\label{fig:a}
\end{figure}

To further demonstrate the physical picture of the transverse sphericity, we plot two typical events with $S_\perp=0$ and $S_\perp=1$ respectively in the $p_xp_y$ plane shown in Fig.~\ref{fig:a}. We can see that $S_\perp=0$ for an event with two jets produced back-to-back as shown in the top panel, and $S_\perp=1$ for an event with multiple jets distributed in the spherically symmetry as shown in the bottom panel. It should be noted that, in the definition of transverse sphericity, it only requires $\lambda_{1}>\lambda_{2}>\lambda_{3}$, therefore transverse sphericity is not necessarily defined in $p_x p_y$ plane. In the following investigation on the jet quenching effect, we only focus on the modification in the transverse plane ($p_x p_y $ plane), so we neglect the $p_{z}$ of the final state jets.

\begin{figure}[!htb]
\centering
\includegraphics[width=9.5cm,height=9.5cm]{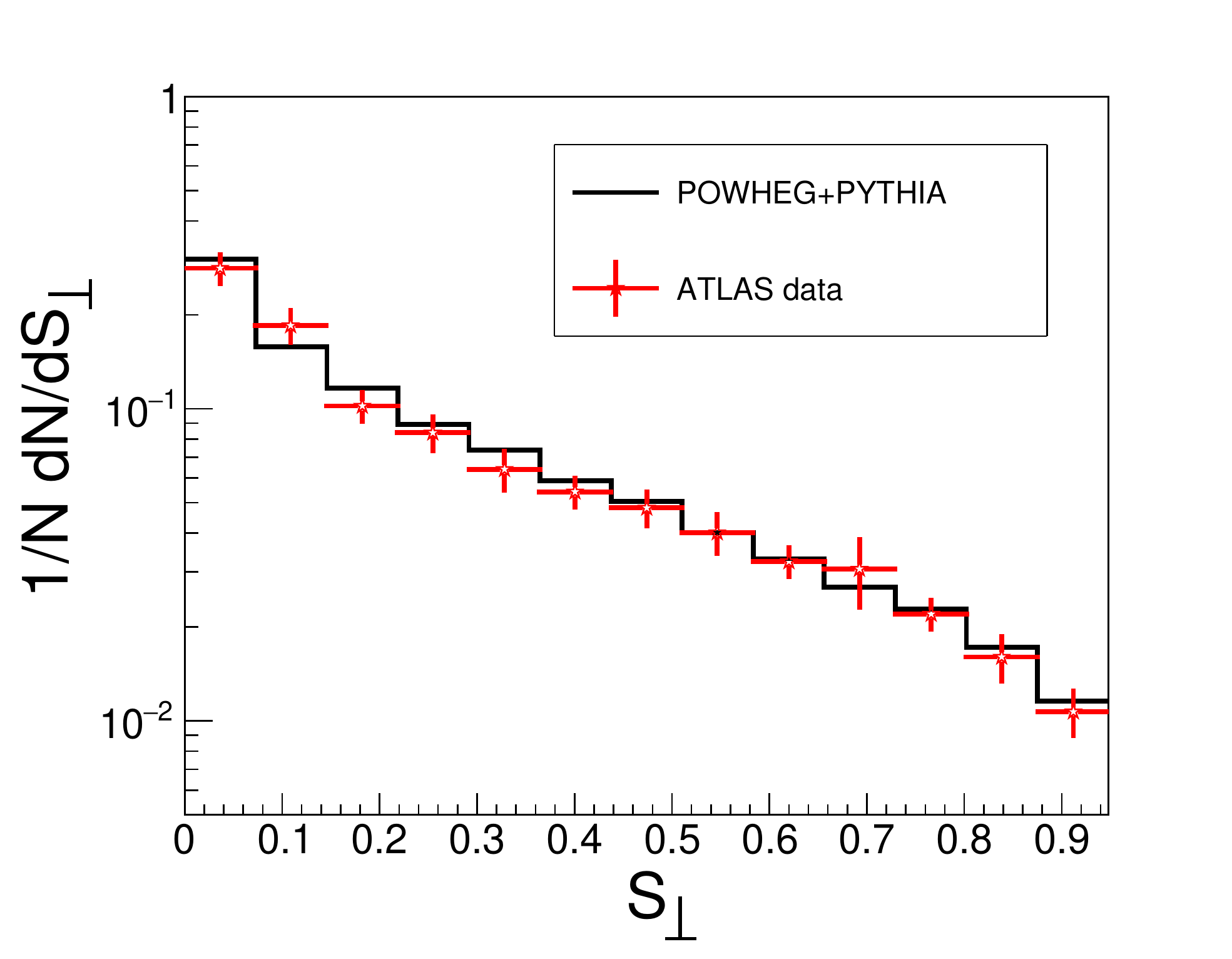} \\
\caption{Event number normalized distribution of transverse sphericity in p+p collisions at $\sqrt{s}=7$~TeV form POWHEG+PYTHIA calculation compared with ATLAS data.}
\label{s-ATLAS}
\end{figure}

To compute event shape observables a satisfactory theoretical description needs both the fixed-order perturbative calculations and the inclusion of the large contributions beyond fixed-order when the event shape observables are very large or
very small~\cite{Banfi:2010xy}. In this paper,
a NLO+PS Monte Carlo event generation framework~\cite{Nason:2004rx, Frixione:2007vw, Alioli:2010xd, Alioli:2010xa} is employed to simulate jet productions in p+p collisions as our baseline. In this framework, NLO matrix elements for QCD dijet process which provided by POWHEG matches with the final state parton showering in PYTHIA6. FastJet package ~\cite{Cacciari:2011ma} is used to reconstruct the final state full jets. Using the framework, one can nicely reproduce a number of  jet production results in p+p collisions~\cite{Alioli:2010xa}. In this manuscript, anti-$k_{T}$ algorithm is used to reconstruct full jets with the radius $R=0.6$, the same kinematic cuts as ATLAS publication~\cite{Aad:2012np} are imposed: the selected events are required to include at least two jets with $(p_{T,1}+p_{T,2})>500$~GeV in central rapidity region $\left|\eta_{1,2}\right|<1.0$, the other jets are required to have $\left|\eta_{j}\right|<1.5$, the lower threshold of the reconstructed jets $p_T $ is $30$~GeV.  We plot the event number normalized production distribution of transverse sphericity in p+p collision at $\sqrt{s_{\rm NN}}=7$~TeV to compare with the ATLAS data shown in Fig.~\ref{s-ATLAS}. We find the simulation using POWHEG+PYTHIA framework can provide fairly good description of the experimental data. Therefore the produced reference p+p events can be served as a reliable baseline for further investigation of their medium modifications.

Before simulating the in-medium evolution of the produced parton jets, we firstly sample the initial positions of the produced jets from Glauber model~\cite{Miller:2007ri}, and directly take the initial events with PS process generated by POWHEG+PYTHIA framework.  A Linear Boltzmann Transport (LBT) model is employed to consider both elastic and inelastic scattering processes of both the initial jet shower partons and the thermal recoil partons with the quasi-particle in the QGP medium~\cite{Wang:2013cia, He:2015pra, Cao:2016gvr}. The elastic scattering process is simulated by the linear Boltzmann transport equation,
\begin{eqnarray}
&p_1\cdot\partial f_a(p_1)=-\int\frac{d^3p_2}{(2\pi)^32E_2}\int\frac{d^3p_3}{(2\pi)^32E_3}\int\frac{d^3p_4}{(2\pi)^32E_4} \nonumber \\
&\frac{1}{2}\sum _{b(c,d)}[f_a(p_1)f_b(p_2)-f_c(p_3)f_d(p_4)]|M_{ab\rightarrow cd}|^2 \nonumber \\
&\times S_2(s,t,u)(2\pi)^4\delta^4(p_1+p_2-p_3-p_4)
 \end{eqnarray}
where $f_{i=a,b,c,d}$ are the phase-space distributions of jet shower partons, $|M_{ab\rightarrow cd}|$ are the corresponding elastic matrix elements which are regulated by a Lorentz-invariant regulation condition $S_2(s,t,u)=\theta(s>2\mu^{2}_{D})\theta(-s+\mu^{2}_{D}\leq t \leq -\mu^{2}_{D})$. $\mu_{D}^{2}=g^{2}T^{2}(N_{c}+N_{f}/2)/3$ is the Debye screening mass. The inelastic scattering is described by the higher twist formalism for induced gluon radiation as~\cite{Guo:2000nz,Zhang:2003yn,Zhang:2003wk},
\begin{equation}
\frac{dN_g}{dxdk_\perp^2 dt}=\frac{2\alpha_sC_AP(x)\hat{q}}{\pi k_\perp^4}\left(\frac{k_\perp^2}{k_\perp^2+x^2M^2}\right)^2\sin^2\left(\frac{t-t_i}{2\tau_f}\right)  .
 \end{equation}
Here $x$ denotes the energy fraction of the radiated gluon relative to parent parton with mass $M$, $k_\perp$ is the transverse momentum. A lower energy cut-off $x_{min}=\mu_{D}/E$ is applied for the emitted gluon in the calculation. $P(x)$ is the splitting function in vacuum, $\tau_f=2Ex(1-x)/(k^2_\perp+x^2M^2)$ is the formation time of the radiated gluons in QGP. The dynamic evolution of bulk medium is given by 3+1D CLVisc hydrodynamical model~\cite{Pang:2012he, Pang:2014ipa} with initial conditions simulated from A Multi-Phase Transport (AMPT) model~\cite{Lin:2004en}. Parameters used in the CLVisc are fixed by reproducing hadron spectra with experimental measurement. In LBT model, $\alpha_s$ is the strong coupling constant which is served as the only one parameter to control the strength of parton-medium interaction. Based on the previous studies~\cite{He:2015pra, Cao:2016gvr, Luo:2018pto, Zhang:2018urd}, we choose $\alpha_s=0.2$ for the following calculations. LBT model has been well tested that could provide nice description of a series of jet quenching measurements, from light and heavy flavor hadrons suppression to single inclusive jets suppression, as well as bosons-jet correlation~\cite{He:2015pra, Cao:2016gvr, Luo:2018pto, Zhang:2018urd}.

\section{Results and Analysis}
\label{sec:results}
We now can simulate the production of the jets in Pb+Pb collisions at $\sqrt{s_{\rm NN}}=5.02$~TeV. In this manuscript, jets are reconstructed with anti-$k_T$ algorithm and radius parameter $R=0.4$ from the candidate partons in an event which are required to have $p_T>500$~MeV. Events are required to have at least two jets: the leading jet $p_T^{\text{leading}}>110$~GeV and the lowest requirement of jet $p_T^{\text{min}}>30$~GeV. All the jets are restricted in the central rapidity region $\vert \eta \vert \leq 2.5$. If the $p_T$ of a jet in the survived event no longer meet the lowest threshold of $p_T^{\text{min}}>30$~GeV after in-medium evolution, then only this jet is discarded. In order to study the hot nuclear alteration of the $S_{\perp}$ distribution, we define the nuclear modification ratio of the normalized distribution as a function of $S_\perp$:
\begin{eqnarray}
R_{AA}^{S_{\perp}}={\dfrac{1}{N_{AA}} \dfrac{dN_{AA}}{dS_{\perp}}}/{\dfrac{1}{N_{pp}} \dfrac{dN_{pp}}{dS_{\perp}}}
\label{eq:ratio}
\end{eqnarray}

\begin{figure}[!htb]
\centering
\vspace{0.in}
\includegraphics[width=9.5cm,height=9.5cm]{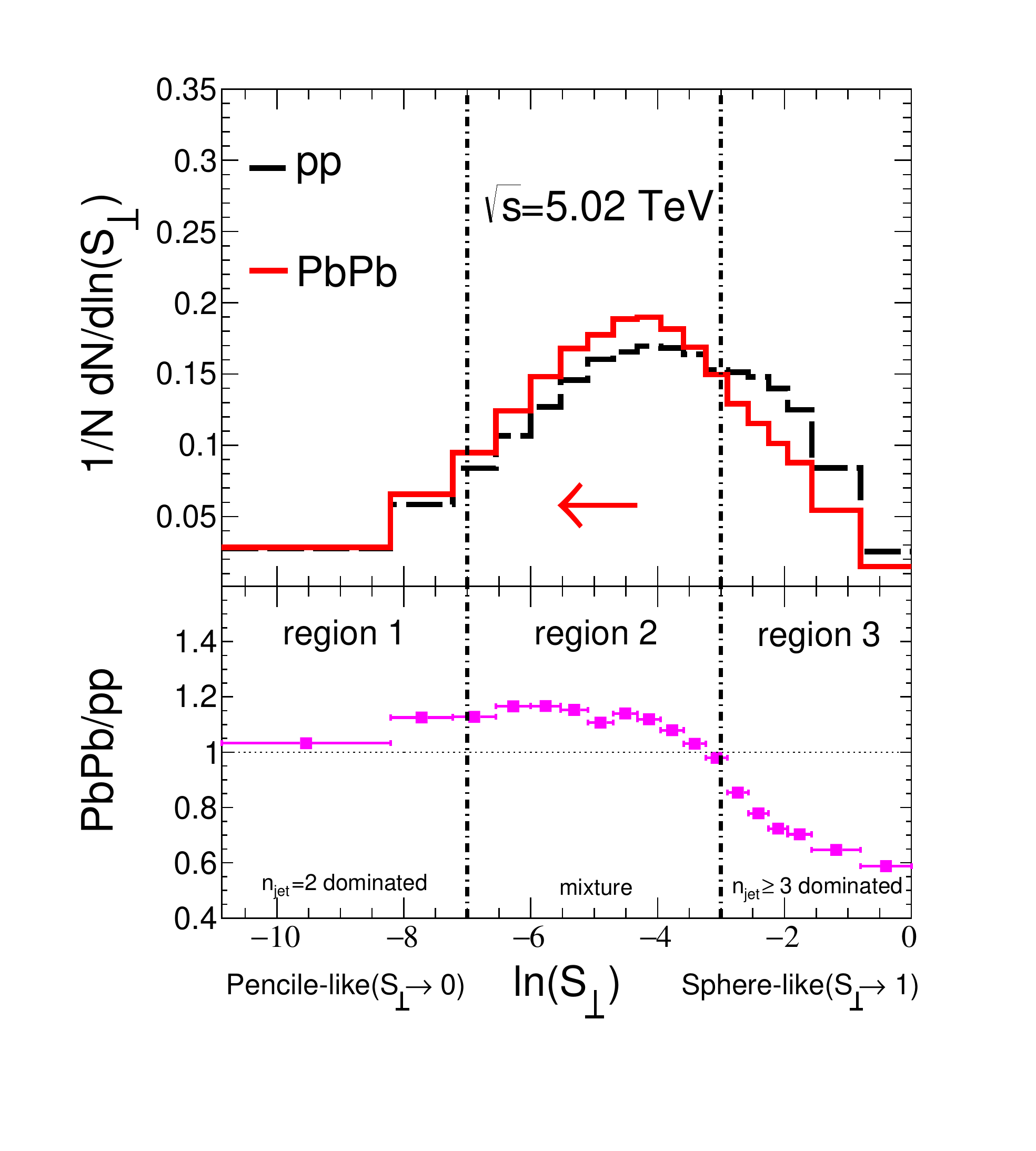} \\
\vspace{-0.6in}
\caption{Top: event normalized transverse sphericity distribution of the total events in p+p and Pb+Pb collisions at $\sqrt{s}=5.02$~TeV;
Bottom: medium modification factor of event normalized $S_{\perp}$ distribution of the total events.}
\label{s-aapp}
\end{figure}

Demonstrated in Fig.~\ref{s-aapp} are the theoretical simulation of the event normalized distribution as functions of $ln (S_\perp)$ in the top plots and its nuclear modification ratio in Pb+Pb collisions at $\sqrt{s}=5.02$~TeV relative to the p+p counterpart in the bottom plots. We find the normalized events distributed widely peaked around $ln(S_\perp)=-4.0$ in p+p collision. The normalized distribution in Pb+Pb collisions is shifted toward smaller $ln (S_\perp)$, therefore leads to an enhancement at smaller $ln (S_\perp)$ regions (region 1 and region 2 in Fig.~\ref{s-aapp}) and suppression at larger $ln (S_\perp)$ region (region 3 in Fig.~\ref{s-aapp}), and the cross point is around $ln(S_\perp)=-3.25$ ($S_\perp=0.03788$). This modification indicates that, there will be more proportion of the survived events being pencil-like after `jet quenching' and leads to less events distributed at lager $S_\perp$. At first glance this result seems to contradict with naive expectation at leading-order that the medium-induced radiative gluon may further spread the energies of jets away from jet axis and then make the energy flow of the system to be more isotropic.

As we mentioned in the Introduction, transverse sphericity was proposed to describe the geometrical properties of an event, it should be relevant to the number of jets and correlation among these jets in an event. Thus, to explore the underlying reasons of medium modification of $S_{\perp}$ distributions, we start the investigation from studying the modifications of $S_{\perp}$ distributions for events with fixed numbers of jets. We separate the events into two categories, events with $2$ jets in the final-state ($n_{\rm jet}=2$) and events with more than $2$ jets in the final state ($n_{\rm jet}\geq3$). Note that the total production fraction of $n_{\rm jet}=2$ is very large ($\sim 57\%$) in p+p collisions at the LHC energies.

\begin{figure}[!htb]
\centering
\vspace{0.in}
\includegraphics[width=9.5cm,height=9.5cm]{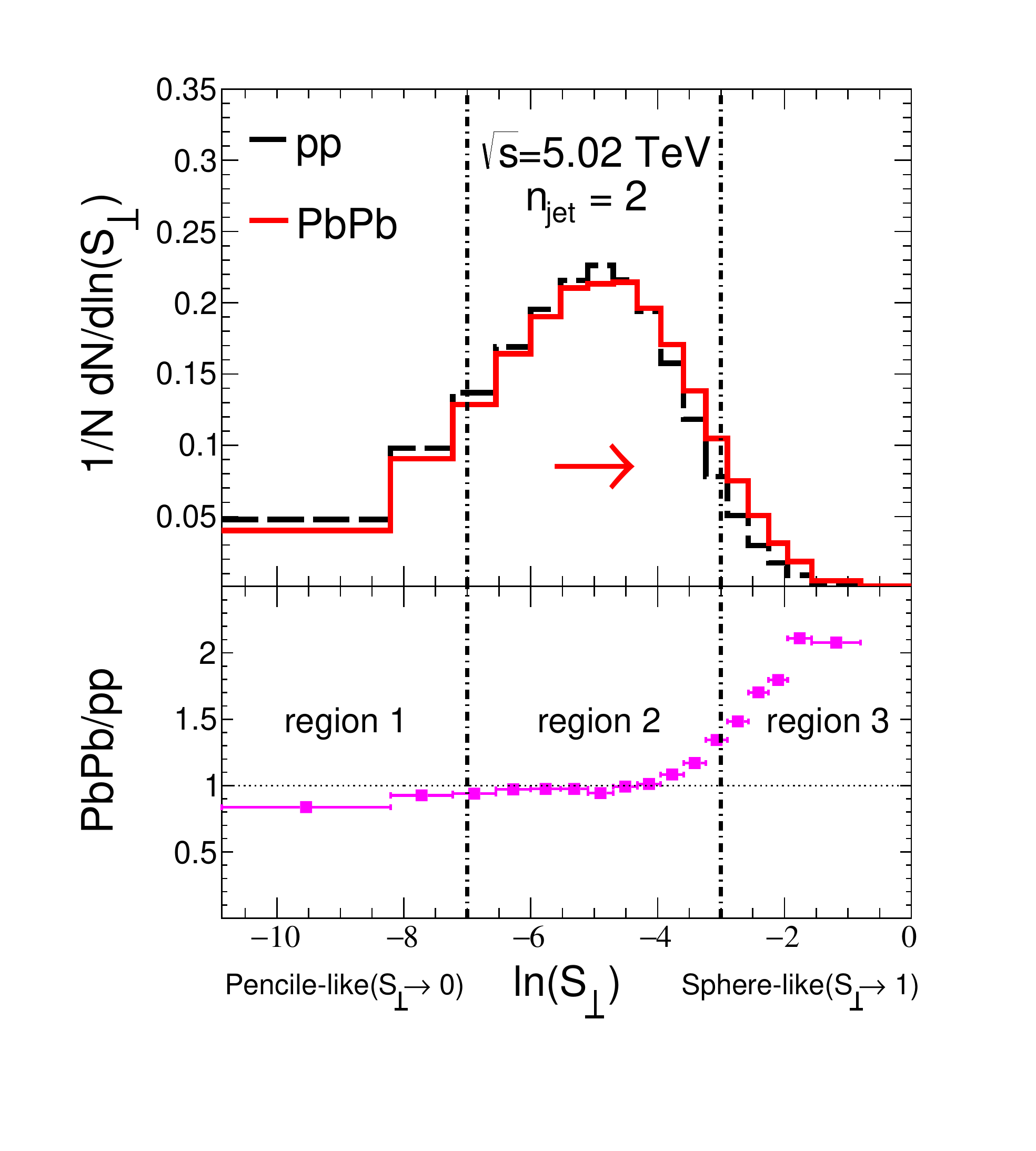}
\vspace{-0.6in}
\caption{Top: event normalized transverse sphericity distribution of $n_\text{jet}=2$ events in p+p and Pb+Pb collisions at $\sqrt{s}=5.02$~TeV;
Bottom: medium modification factor of event normalized $S_{\perp}$ distribution of $n_\text{jet}=2$ events.}
\label{s2-aapp}
\end{figure}

Then let us look at the medium modification effect of event normalized $S_\perp$ distribution when only considering events with $n_\text{jet}=2$. We plot in Fig.~\ref{s2-aapp} the event normalized transverse sphericity distribution of $n_\text{jet}=2$ events in p+p and Pb+Pb collisions at $\sqrt{s}=5.02$~TeV and also its medium modification factor in the bottom panel. We find $n_\text{jet}=2$ events are centered around $ln(S_{\perp})=-5$ in both p+p and Pb+Pb collisions. The event normalized $S_\perp$ distribution is moderately suppressed in $ln(S_\perp)\in(-10.87,-7)$ region (region 1), but largely enhanced in $ln(S_\perp)\in(-3,0)$ region (region 3) in Fig.~\ref{s2-aapp} . In the region with $ln(S_\perp)\in(-7,-3)$ (region 2) , the modification is minor. The medium modification to the $n_\text{jet}=2$ events is to reduce the proportion of events that are originally distributed back-to-back, and therefore enhance the proportion distributed at larger $S_\perp$.  In a word, $n_\text{jet}=2$ events are shifted towards sphere-like in Pb+Pb collisions compared with that in p+p. The trend of such distribution shifting is  in agreement with what we may expect for the jet quenching effect on the modification of the $S_\perp$ distribution at leading-order estimation, that medium-induced gluon radiation may carry way energies from hard partons and make the two jets not exactly back-to-back and the energy flow more isotropic.  To benefit further discussion, we also define $R^{n_\text{jet}=2}_{AA}$ to denote the ratio of the normalized $S_\perp$ distribution in A+A and p+p (plotted in the bottom panel of Fig.~\ref{s2-aapp}):

\begin{eqnarray}
R^{n_\text{jet}=2}_{AA}={\dfrac{1}{N^{n_\text{jet}=2}_{AA}} \dfrac{dN^{n_\text{jet}=2}_{AA}}{dS_{\perp}}}/{\dfrac{1}{N^{n_\text{jet}=2}_{pp}} \dfrac{dN^{n_\text{jet}=2}_{pp}}{dS_{\perp}}}
\label{eq:ratio2}
\end{eqnarray}

\begin{figure}[!htb]
\centering
\vspace{0.in}
\includegraphics[width=9.5cm,height=9.5cm]{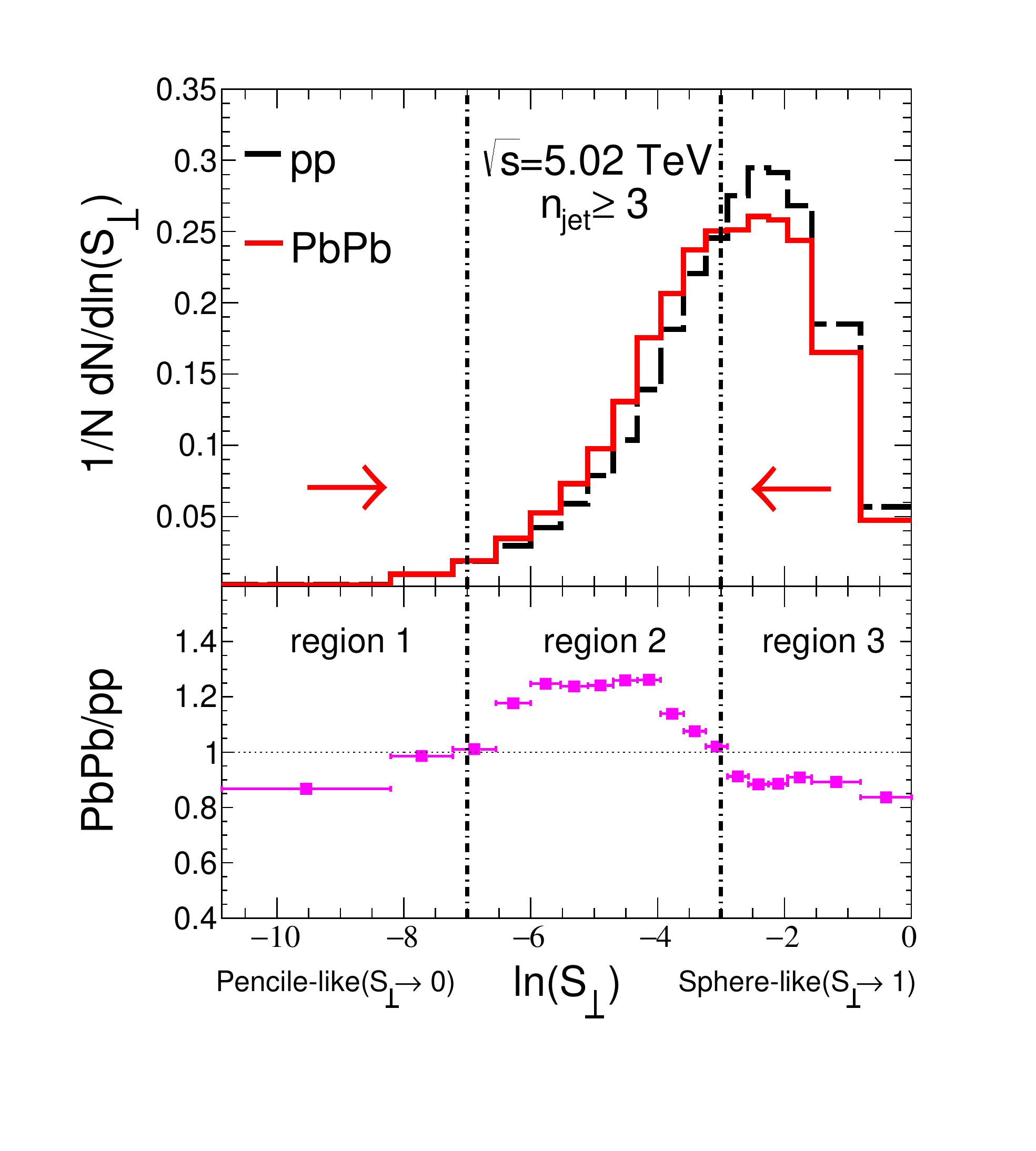} \\
\vspace{-0.6in}
\caption{Top: event normalized transverse sphericity distribution of $n_\text{jet}\geq3$ events in p+p and Pb+Pb collisions at $\sqrt{s}=5.02$~TeV;
Bottom: medium modification factor of event normalized $S_{\perp}$ distribution of $n_\text{jet}\geq3$ events.}
\label{s3-aapp}
\end{figure}

We further plot in Fig.~\ref{s3-aapp} the event normalized transverse sphericity $S_{\perp}$ distribution for events with jets number $n_\text{jet}\geq3$ in p+p and Pb+Pb collisions at $\sqrt{s}=5.02$~TeV as well as its nuclear modification factor defined as:
\begin{eqnarray}
R^{n_\text{jet}\geq3}_{AA}={\dfrac{1}{N^{n_\text{jet}\geq3}_{AA}} \dfrac{dN^{n_\text{jet}\geq3}_{AA}}{dS_{\perp}}}/{\dfrac{1}{N^{n_\text{jet}\geq3}_{pp}} \dfrac{dN^{n_\text{jet}\geq3}_{pp}}{dS_{\perp}}}
\label{eq:ratio3}
\end{eqnarray}

We find there is an enhancement at the $ln(S_\perp)\in(-7,-3)$ bins (region.2) and suppression in the other $S_{\perp}$ region. It is because in the region $ln(S_\perp)\in(-10.87,-7)$, jets are relatively balanced. After jet quenching, this kind of balance will be broken and leads to a higher value of $S_{\perp}$.
Besides, in $n_\text{jet}\geq3$ events, jets are relatively softer as compared to $n_\text{jet}=2$ events, thus jets are earlier to lose energy due to jet quenching effect and then fall below the lower threshold of the jet selection, which makes $n_\text{jet}$ of such event to decrease (referred as {\it jet number reduction effect}), for instance, from $n_\text{jet}=4$ to $n_\text{jet}=3$. The similar jet number reduction effect of $n_\text{jet}=3$ events to $n_\text{jet}=2$ events has also been revealed in~[60], which showed that initial 3-jet like events in p+p collisions, when quenched, are more similar to (imbalance) di-jet events in Pb+Pb collisions~\cite{Apolinario:2012cg}.
As a consequence of jet number reduction effect, the proportional distribution at lager $S_{\perp}$ will be reduced
significantly.  The effect of jet number alteration due to the jet quenching may cause the multi-jet events more pencil-like (or jetty), because there are less jets in the final-state of events. We have plotted the nuclear modification factor of $S_{\perp}$ when only considering $n_\text{jet}\geq3$ events in the bottom panel of Fig.~\ref{s3-aapp}.

Now we consider the transverse sphericity distribution by including both events with $n_\text{jet}=2$ and events $n_\text{jet}\geq3$. In p+p collisions,
the total event normalized $S_{\perp}$ distribution can then be expressed as:

\begin{eqnarray}
{\dfrac{1}{N_{pp}} \dfrac{dN_{pp}}{dS_{\perp}}}={\dfrac{1}{N_{pp}} \dfrac{dN_{pp}^{n_\text{jet}=2}}{dS_{\perp}}}+{\dfrac{1}{N_{pp}} \dfrac{dN_{pp}^{n_\text{jet}\geq3}}{dS_{\perp}}}
\label{eq:frac}
\end{eqnarray}

\begin{figure}[!htb]
\centering
\includegraphics[width=8.5cm,height=8.5cm]{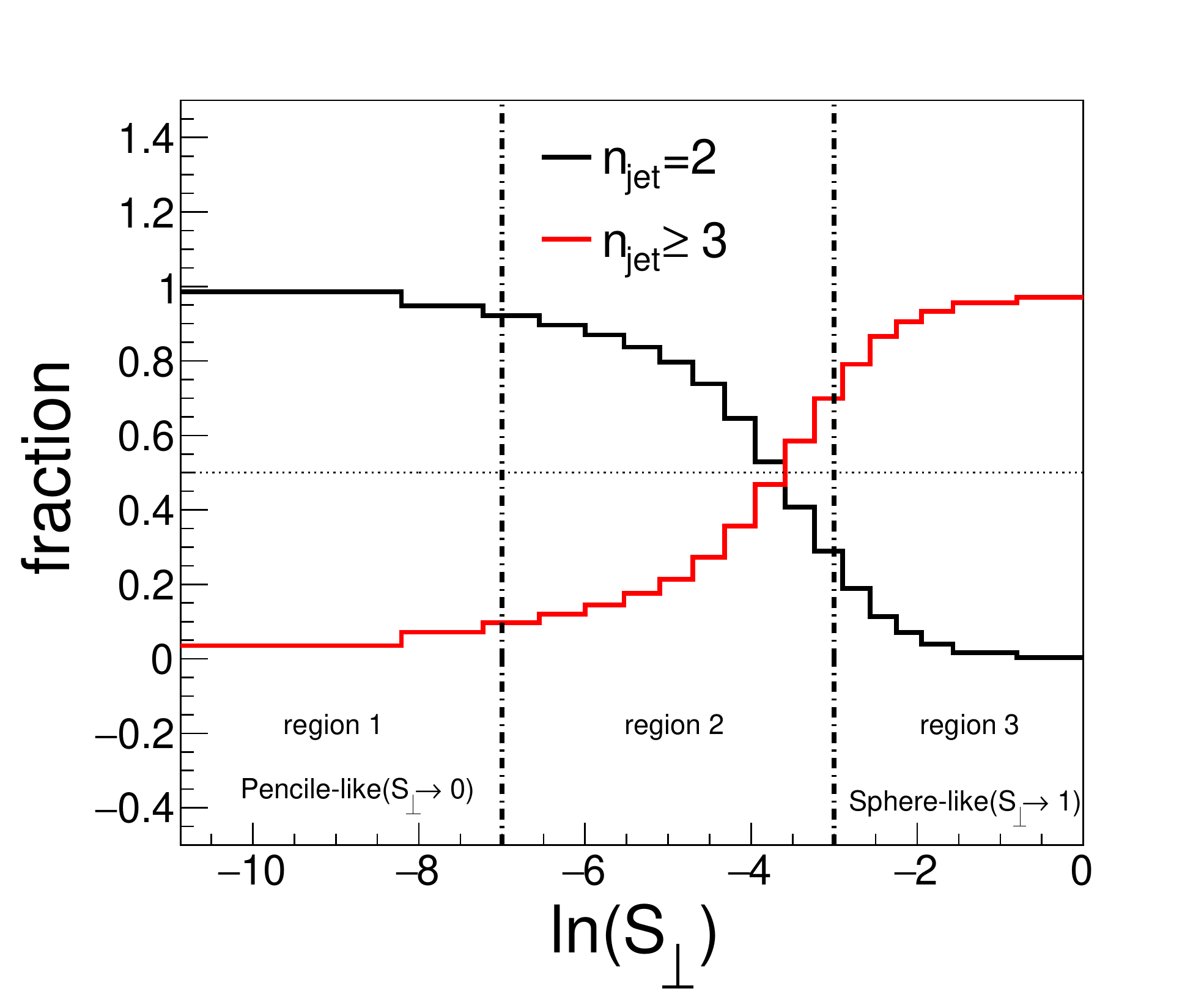} \\
\caption{Relative contribution fraction of
total event number normalized transverse sphericity distribution of $n_\text{jet}=2$, $n_\text{jet}\geq3$ events in p+p collisions at $\sqrt{s}=5.02$~TeV.}
\label{s_njets}
\end{figure}

We plot in Fig.~\ref{s_njets} the production fractions as functions of $S_{\perp}$ for $n_\text{jet}=2$ events and $n_\text{jet}\geq3$ events respectively in p+p collision at $\sqrt{s_{\rm NN}}=5.02$~TeV. From the rapid fall of the two jets events ($n_\text{jet}=2$) contribution shown in the black line, we find the two jets events will give more contribution at the lowest bin $ln(S_\perp)\in(-10.87,-7)$ than $n_\text{jet}\geq3$ events, we refer the region $ln(S_\perp)\in(-10.87,-7)$ as $n_\text{jet}=2$ dominated area. While $ln(S_\perp)\in(-3,0)$ as $n_\text{jet}\geq3$ dominated area. The intermediate region $ln(S_\perp)\in(-7,-3)$ is $n_\text{jet}=2$ and $n_\text{jet}\geq3$ mixture area.

\begin{figure}[!htb]
\centering
\vspace{0.in}
\includegraphics[width=11cm]{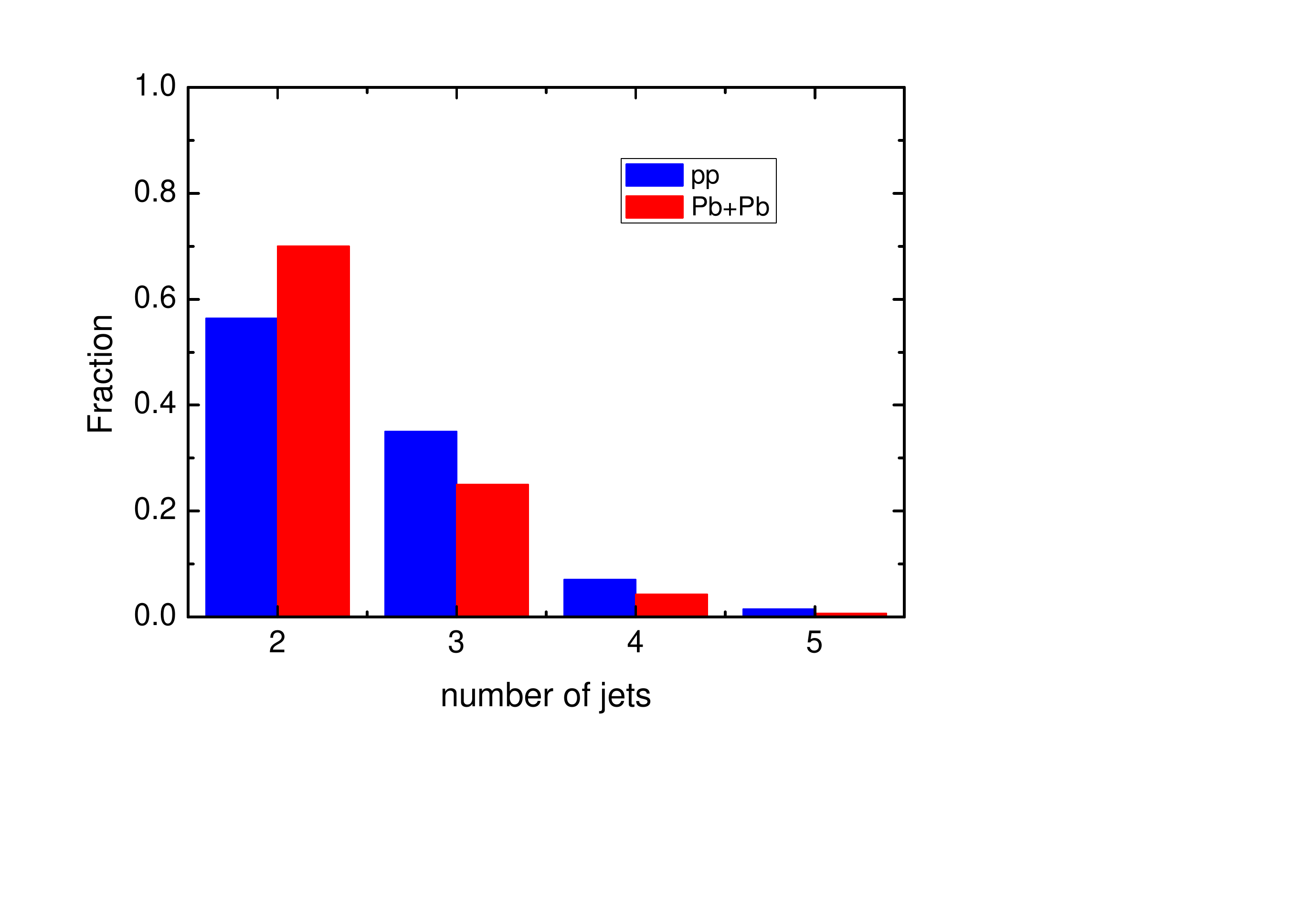} \\
\vspace{-0.6in}
\caption{The production proportion of $n_\text{jet}=2$ events and $n_\text{jet}\geq3$ events in p+p and Pb+Pb collisions at $\sqrt{s}=5.02$~TeV.}
\label{number}
\end{figure}

The jet number reduction effect will also cause the shifting between of the relative production proportion of $n_\text{jet}=2$ events and $n_\text{jet}\geq3$ events, which is important to understand the medium modification of the event normalized $S_\perp$ distribution. We demonstrate in Fig.~\ref{number} the total production proportion of $n_\text{jet}=2$ events and $n_\text{jet}\geq3$ events in p+p and Pb+Pb collisions at $\sqrt{s}=5.02$~TeV. It shows that the proportion of the $n_\text{jet}=2$ events in p+p collision $f^{n_\text{jet}=2}_{pp}$ is $57\%$, and it will increase to $f^{n_\text{jet}=2}_{AA}=70\%$ in Pb+Pb collisions. That is why we refer the variation of total production  proportion of $n_\text{jet}\geq3$ events in Pb+Pb collisions as jet number reduction effect.

Using the notation mentioned in the above discussion, we can easily express the nuclear modification factor of transverse distributions in the Eq.~\ref{eq:ratio} as:
\begin{widetext}
\begin{eqnarray}
     R_{AA}^{S_{\perp}}&=& \frac{R^{n_\text{jet}=2}_{AA}\dfrac{dN^{n_\text{jet}=2}_{pp}}{dS_{\perp}} \dfrac{K^{n_\text{jet}=2}_{AA}}{K^{n_\text{jet}=2}_{pp}}+R^{n_\text{jet}\geq3}_{AA}\dfrac{dN^{n_\text{jet}\geq3}_{pp}}{dS_{\perp}} \dfrac{(1-K^{n_\text{jet}=2}_{AA})}{(1-K^{n_\text{jet}=2}_{pp})}}{\dfrac{dN^{n_\text{jet}=2}_{pp}}{dS_{\perp}}+
\dfrac{dN^{n_\text{jet}\geq3}_{pp}}{dS_{\perp}} }  \nonumber \\
     &=& R^{n_\text{jet}=2}_{AA}f^{n_\text{jet}=2}_{pp}(S_\perp)\dfrac{K^{n_\text{jet}=2}_{AA}}{K^{n_\text{jet}=2}_{pp}}+ R^{n_\text{jet}\geq3}_{AA}(1- f^{n_\text{jet}=2}_{pp}(S_\perp))\dfrac{(1-K^{n_\text{jet}=2}_{AA})}{(1-K^{n_\text{jet}=2}_{pp})}
\label{eq:r2r3}
\end{eqnarray}
\end{widetext}

Now we are ready to analyze the medium modification of $S_{\perp}$ distributions at three $S_{\perp}$ regions as we shown in Fig.~\ref{s-aapp}.
From the plots in Fig.~\ref{number}, we can estimate the $\dfrac{K^{n_\text{jet}=2}_{AA}}{K^{n_\text{jet}=2}_{pp}}\approx1.228$ and $\dfrac{(1-K^{n_\text{jet}=2}_{AA})}{(1-K^{n_\text{jet}=2}_{pp})}\approx 0.7$.
First, we look at $n_\text{jet}=2$ events dominated region (region 1). Since in this region $f^{n_\text{jet}=2}_{pp}\approx 1$ (see Fig.~\ref{s_njets}), we can only consider the first term to calculate $R_{AA}^{S_{\perp}}$ in
Eq.~\ref{eq:r2r3}. Therefore in this region, we have the total medium modification of $S_{\perp}$ distributions $R_{AA}^{S_{\perp}}\approx R^{n_\text{jet}}_{AA}\dfrac{K^{n_\text{jet}=2}_{AA}}{K^{n_\text{jet}=2}_{pp}}$ with $R^{n_\text{jet}=2}\approx0.85$ (see Fig.~\ref{s2-aapp}).
We know $R_{AA}^{S_{\perp}}\approx 0.85\cdot 1.228 = 1.04$ in Fig.~\ref{s-aapp} in $n_\text{jet}=2$ dominated region.
This is the region where jet number reduction effect playing an essential role to overcome the trend of $R_{AA}^{S_{\perp}}$ from suppression for $n_\text{jet}=2$ events,
toward enhancement for total events.
Next, we look at the nuclear modification ratio $R_{AA}^{S_{\perp}}$ at $n_\text{jet}\geq3$ dominated area (region 3). In this region, since $f^{n_\text{jet}\geq3}_{pp}\approx 1$, we could neglect the contributions from $n_\text{jet}=2$ events even though the distributions of $n_\text{jet}=2$ events are largely enhanced. It is shown that the $S_{\perp}$ distributions of $n_\text{jet}\geq3$ events are moderate
suppressed ($R^{n_\text{jet}\geq3}_{AA}\approx 0.87$ in the second term of Eq.~\ref{eq:r2r3}  ). By the effect of the jet number reduction, we have $R_{AA}^{S_{\perp}}\approx R^{n\geq3}_{AA}\cdot \dfrac{(1-K^{n_\text{jet}=2}_{AA})}{(1-K^{n_\text{jet}=2}_{pp})}\approx 0.6$ at $n_\text{jet}\geq3$ dominated area shown in Fig.~\ref{s-aapp}.
At the mixed events region (region 2),  the modifications to the $n_\text{jet}=2$ events are relatively small which could be observed in Fig.~\ref{s2-aapp} and the enhancement to the $n_\text{jet}\geq3$ events will lead to the final enhancement shown the Fig.~\ref{s-aapp}.

\begin{figure}[!htb]
\centering
\includegraphics[width=7cm,height=7cm]{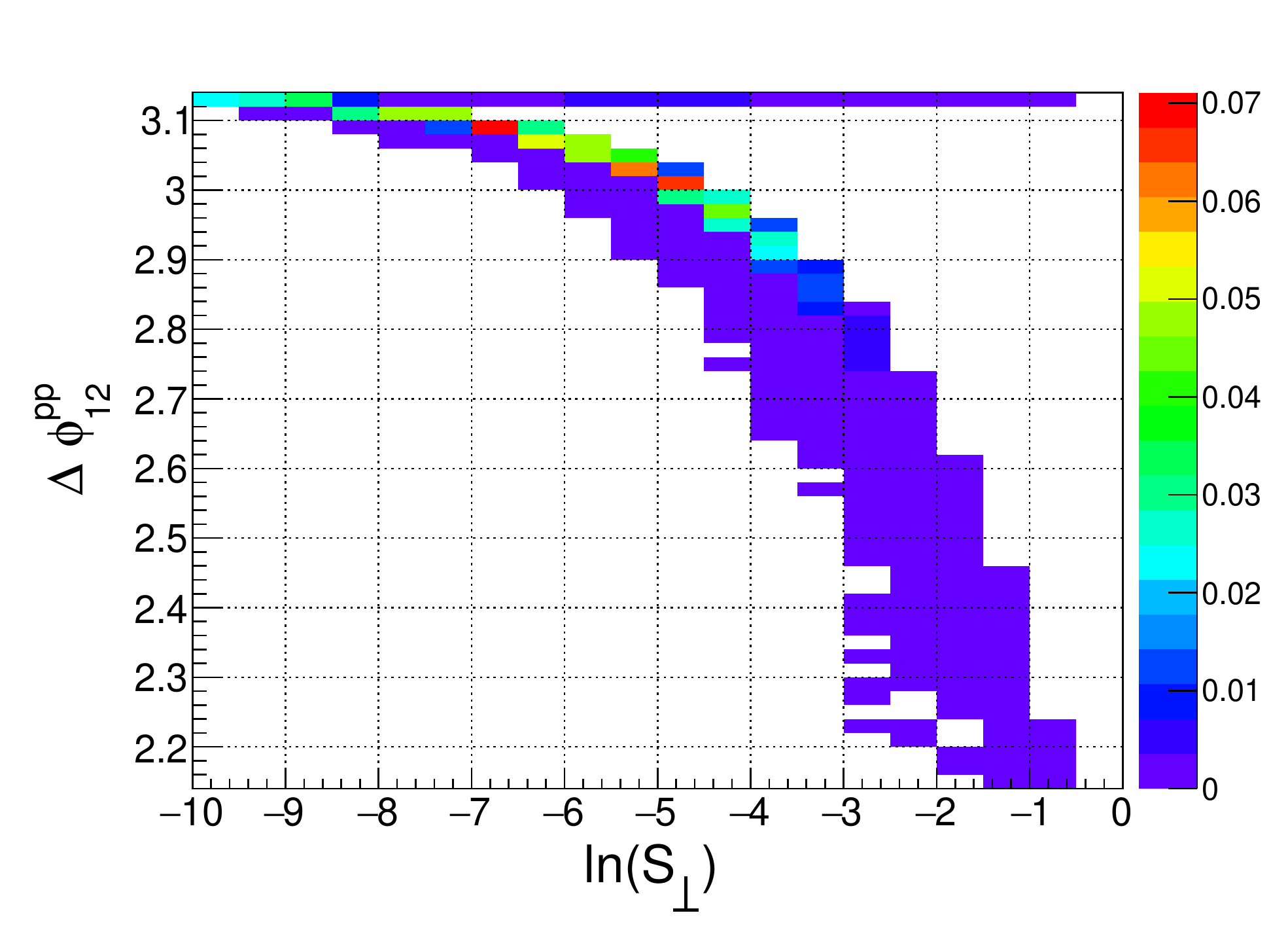}
\includegraphics[width=7cm,height=7cm]{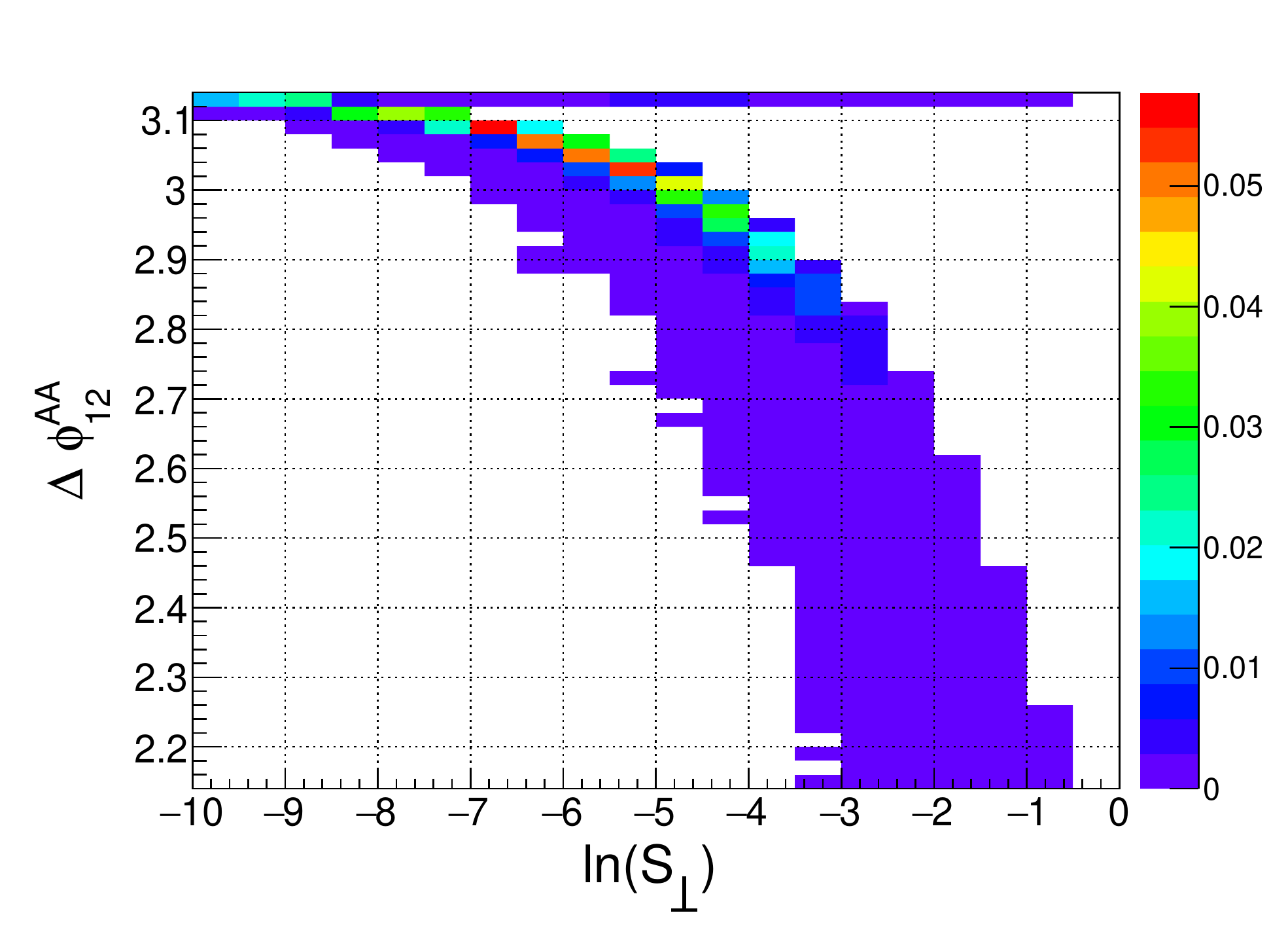}\\
\caption{Correlation between $S_{\perp}$ and $\Delta \phi_{12}$ for $n_\text{jet}=2$ events in p+p (top) and Pb+Pb (bottom) collisions at $\sqrt{s}=5.02$~TeV.}
\label{s_phi_2}
\end{figure}

Finally, to explore the underlying factors that could be responsible for the medium modifications of the $S_{\perp}$ distributions in events with identical jet numbers, we conduct correlation studies in $n_\text{jet}=2$ events, since $n_\text{jet}=2$ events are clear in momentum space. The azimuth angle ($\Delta\phi$) and transverse momentum balance ($x_{J}$) between the two jets which defined as

\begin{eqnarray}
\Delta \phi_{12}=\left|\phi_{1}-\phi_{2}\right|,
\label{eq:phi}
\end{eqnarray}

\begin{eqnarray}
x_{J}=\frac{p_{T,2}}{p_{T,1}}
\label{eq:xj}
\end{eqnarray}
respectively are the two observables related to $S_\perp$.

\begin{figure}[!htb]
\centering
\includegraphics[width=7cm,height=7cm]{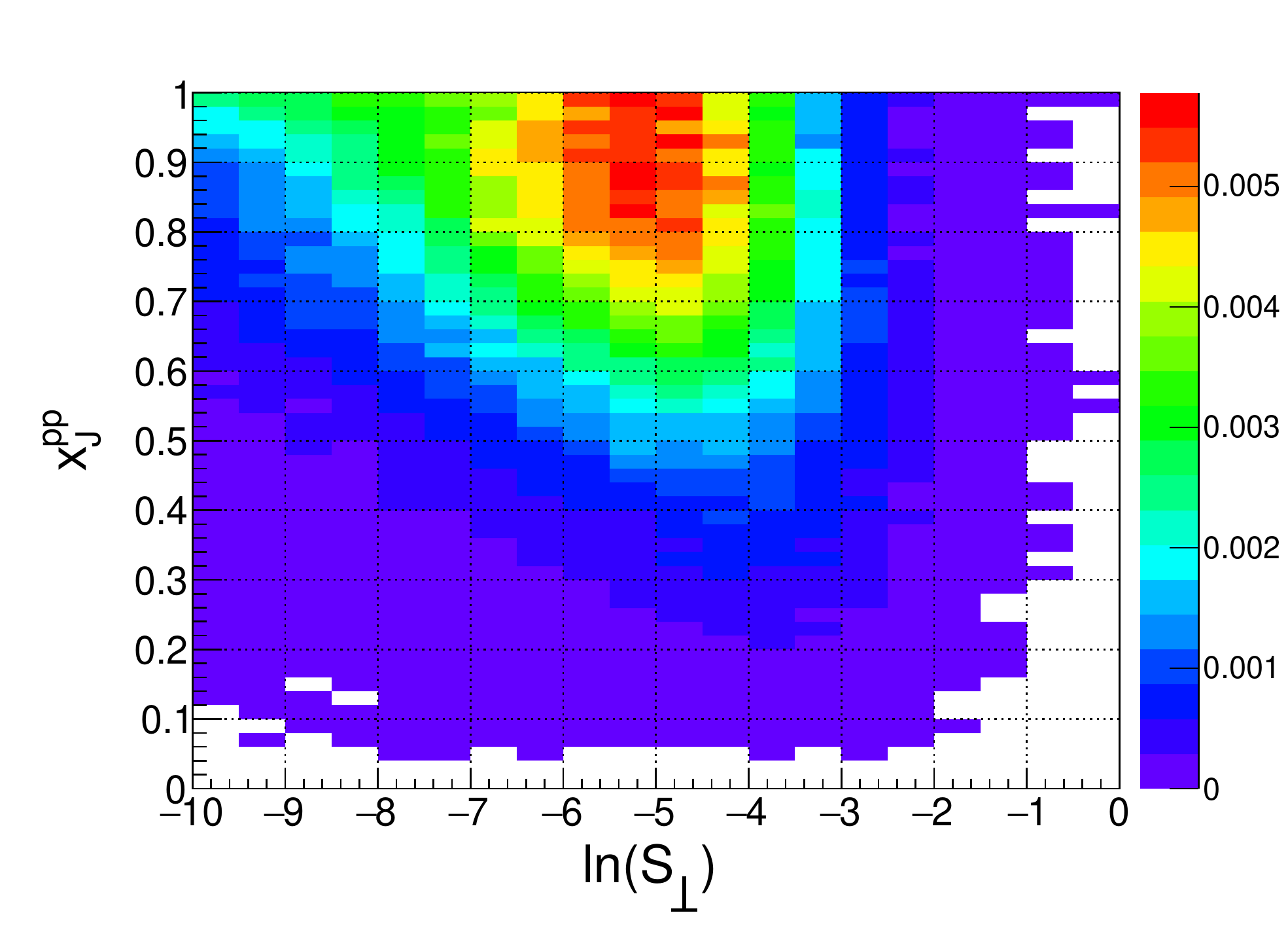}
\includegraphics[width=7cm,height=7cm]{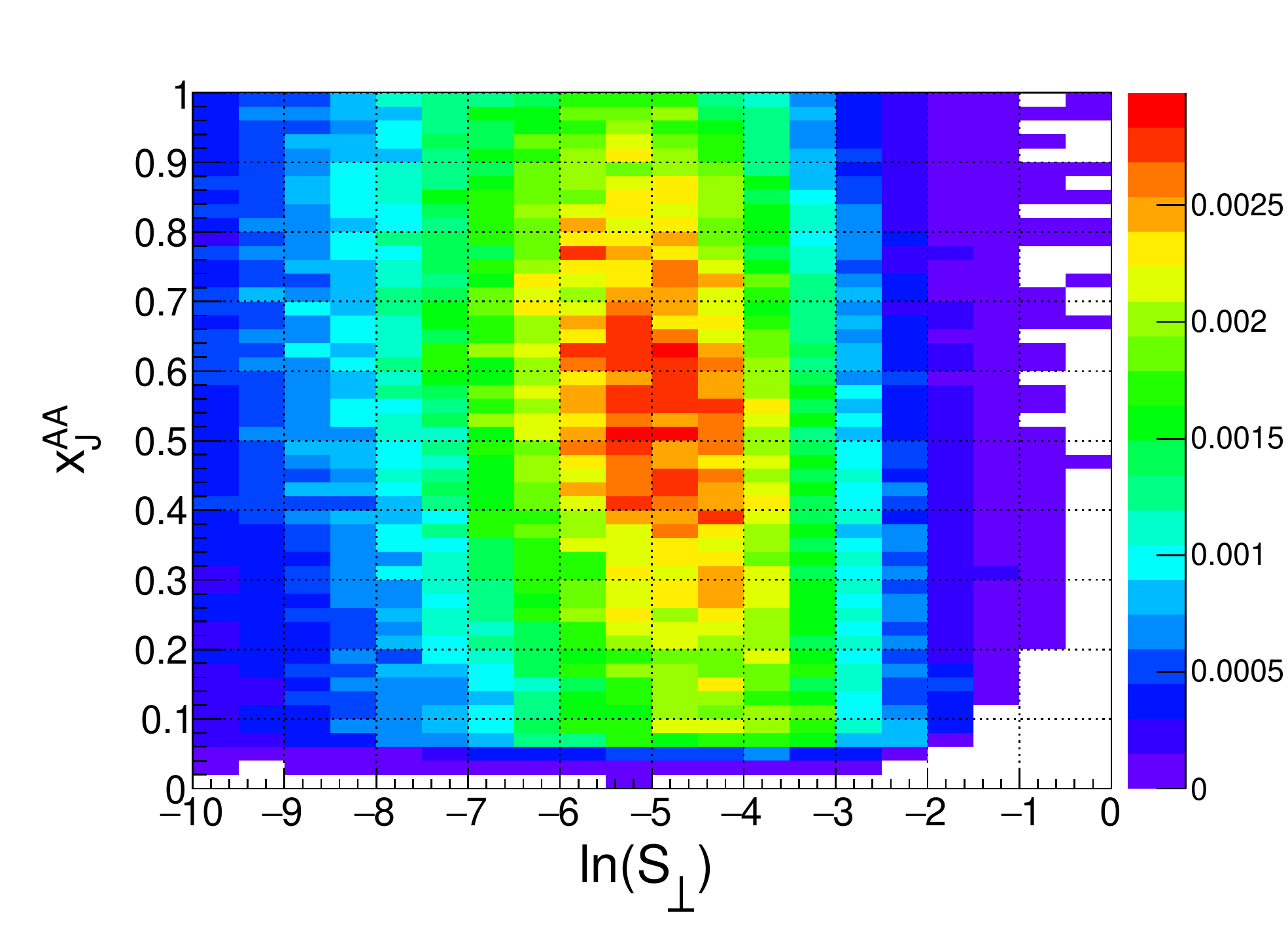}\\
\caption{Correlation between $S_{\perp}$ and $x_{J}$ for $n_\text{jet}=2$ events in p+p (top) and Pb+Pb (bottom) collisions at $\sqrt{s}=5.02$~TeV.}
\label{s_xj_2}
\end{figure}

We plot the correlation between $ln(S_{\perp})$ and $\Delta \phi_{12}$ for $n_\text{jet}=2$ events both in p+p and Pb+Pb collisions at $\sqrt{s}=5.02$~TeV in Fig.~\ref{s_phi_2}. The comparison between the upper and bottom panel shows a very mild correlation variation in A+A and p+p collisions even we find the jet quenching effect will slightly suppress the event number normalized $\Delta \phi_{12}$ distribution by $\sim10\%$ in the back-to-back region ($\Delta \phi_{12}\sim \pi$) while largely enhance in the $\Delta \phi_{12}\sim2.14-2.9$ region by $\sim300\%$ in the calculation which lead the $n_\text{jet}=2$ events to be more isotropic. Noticing the medium modification of $ln(S_{\perp})$ and $\Delta \phi_{12}$ correlation is not very visible, we also plot the correlation between $ln(S_{\perp})$ and $x_{J}$ both in p+p and Pb+Pb collisions in Fig.~\ref{s_xj_2}.  In p+p collisions, events are centered around the region where the transverse momentum of the two jets  are balance ($x_{J}\in(0.8,1)$). After the in-medium evolution, events are centered around smaller  $x_{J}$ region ($x_{J}\in(0.4,0.7)$)  and will naturally lead $n_\text{jet}=2$ events to be more isotropic.  We can conclude from the two comparisons that the medium modification of $S_{\perp}$ in $n_\text{jet}=2$ events is more sensitive to the downshifting of the $x_{J}$ distribution.

\section{Summary}
\label{sec:summary}
In this paper, we present the first theoretical results of the medium modification of transverse sphericity distribution due to jet quenching effect in heavy-ion collisions at large momentum transfer. In our investigation, POWHEG+PYTHIA is employed to provide the p+p baseline up to the next-to-leading order (NLO) accuracy with resummation by matched parton shower. The Linear Boltzmann Transport (LBT) model of the parton energy loss is implemented to simulate the in-medium evolution of jets. We calculate the event normalized medium modification factor as a function of transverse sphericity distributions in the overall region. An enhancement at small transverse
sphericity region and a suppression at large transverse sphericity region are observed in
Pb+Pb collisions compared to their p+p references, which implies medium modification of the event shape is towards pencil-like relative to that in p+p.

To further explore the underlying reasons of the medium alteration of transverse sphericity
distribution, we categorizes the events into two types: $n_\text{jet}=2$ and $n_\text{jet}\geq3$ events. Shifting of nuclear modification for $n_\text{jet}=2$ events is towards sphere-like, and that for $n_\text{jet}\geq3$ events towards pencil-like.
Moreover, we found the fraction of $n_\text{jet}=2$ events will be enhanced after jet quenching in Pb+Pb collisions compared with that in p+p. This will lead to the enhancement of events distributions in low $S_{\perp}$ region because $n_\text{jet}=2$ events give main contribution of the distributions in $ln(S_{\perp})\in(-10.87,-7)$ bins. While in larger $ln(S_{\perp})$ region, events are dominated by $n_\text{jet}\geq3$ in which jets usually have relatively smaller energies and may fall off the jet selection kinematic cut after their energy loss in the medium. This jet number reduction effect will lead to the suppression at large $S_{\perp}$ region for $n_\text{jet}\geq3$ events. The overall trend of the global geometric pattern of events in Pb+Pb is more pencil-like (or jetty) relative to that in p+p, because jet number reduction effect is more pronounced.


{\bf Acknowledgments:}  The authors would like to thank H Zhang, T Luo, P Ru, G Ma, J Yan and S Wang for helpful discussions. This research is supported by Natural Science Foundation of China with Project No. 11935007, 11805167.

\end{document}